\documentclass{PoS}

\usepackage{calc}
\usepackage{graphicx}
\usepackage{amsmath}
\usepackage{amssymb}
\usepackage{relsize}
\usepackage{multirow}
\usepackage{bm}
\usepackage{cite}

\usepackage{graphicx}

\newcommand{\be}{\begin{equation}}
\newcommand{\ee}{\end{equation}}
\newcommand{\ba}{\begin{eqnarray}}
\newcommand{\ea}{\end{eqnarray}}
\newcommand{\bi}{\begin{itemize}}
\newcommand{\ei}{\end{itemize}}

\newcommand{\la}{\label}

\newcommand{\txts}{\textstyle}

\title{Adler function and hadronic vacuum polarization from lattice vector correlation functions in the time-momentum representation}

\ShortTitle{Adler function and hadronic vacuum polarization from lattice vector correlators}

\author{\speaker{Anthony Francis}$^3$, Benjamin J\"ager$^2$, Harvey~B.~Meyer$^{1,2,3}$ and Hartmut Wittig$^{1,2,3}$\\
        PRISMA Cluster of Excellence$^1$,
        Institut f\"ur Kernphysik$^2$ and Helmholtz~Institut~Mainz$^3$,
        Johannes Gutenberg-Universit\"at Mainz,
        D-55099 Mainz, Germany\\
        E-mail: \email{francis@kph.uni-mainz.de},
        \email{jaeger@kph.uni-mainz.de},\\
        \email{meyerh@kph.uni-mainz.de}, 
        \email{wittig@kph.uni-mainz.de}
        
        }

\abstract{We study a representation of the hadronic vacuum polarization based on the time-momentum representation of the vector correlator. This representation suggests a way to compute the hadronic vacuum polarization and the associated Adler function for any value of virtuality, irrespective of the flavor structure of the current. We present results on both of these phenomenologically important functions, derived from local-conserved two-point lattice vector correlation functions, computed on a subset of light two-flavor ensembles made available to us through the CLS effort.}

\FullConference{31st International Symposium on Lattice Field Theory LATTICE 2013\\
                 July 29 - August 3, 2013\\
                 Mainz, Germany}

\begin{document}

\section{Introduction}
The hadronic vacuum polarization $\Pi(Q^2)$ is of great importance in precision tests of the Standard Model of particle physics. It enters, for instance, the running of the QED coupling constant. Additionally, it currently represents the dominant uncertainty in the Standard Model prediction of the anomalous magnetic moment of the muon $a_\mu^{\rm HLO}$.\\
Studies of $a_\mu^{\rm HLO}$ on the lattice have had to deal with the fact that only a finite set of virtualities $Q^2$ were available and an extrapolation of $\Pi(Q^2)\rightarrow\Pi(0)$ had to be performed in order to form the difference $\widehat\Pi(Q^2)$ \cite{gen1, gen2, gen3, gen4, gen5, gen6, gen7, gen8, gen9}. 
Here, we report on a new way of computing $\widehat\Pi(Q^2)$ as well as the Adler function $D(Q^2)$ at any virtuality directly from the time-momentum representation vector correlation function on the lattice, as it was presented before in \cite{Francis:2013fzp,Bernecker:2011gh} and in a somewhat similar spirit in \cite{ETM}.

\section{$\Pi(Q^2)$ and $a_\mu^{\rm HLO}$ in Euclidean space}
On a Euclidean lattice the vacuum polarization tensor can be defined as the four dimensional Fourier transform of the vector current-current correlation function:
\be\la{eqn:fourcorr}
\Pi_{\mu\nu}(Q) \equiv \int d^4x \, e^{iQ\cdot x} \langle j_\mu(x) j_\nu(0)\rangle.
\ee
Here $O(4)$ invariance and current conservation imply the tensor structure
\be
\la{eqn:tensorstruct}
\Pi_{\mu\nu}(Q) =\big(Q_\mu Q_\nu -\delta_{\mu\nu}Q^2\big) \Pi(Q^2).
\ee
    
The lowest order hadronic contribution to the anomalous magnetic moment of the muon $a_\mu^{\rm HLO}$ can be related to $\Pi(Q^2)$ through, 
\be\la{eqn:gminus}
a_\mu^{\rm HLO} = \Big(\frac{\alpha}{\pi}\Big)^2\int dQ^2 K_E(Q^2,m_\mu) \widehat{\Pi}(Q^2).
\ee
here, the kernel is known from QED and the hadronic part $\widehat{\Pi}(Q^2)=4\pi^2( \Pi(Q^2) - \Pi(0) )$ can be determined on the lattice.
However, following this recipe and computing $\widehat{\Pi}(Q^2)$ via (\ref{eqn:fourcorr}) on the lattice one is faced with the problem that the intercept $\Pi(Q^2=0)$ is not directly available and it has to be estimated using an extrapolation procedure. In addition the integrand of (\ref{eqn:gminus}) is strongly peaked around the lepton's mass, and with the muon mass at $m_\mu\simeq105.65$MeV \cite{PDG}, this is generally below the lattice momentum resolution.

\section{A new representation for $\widehat\Pi(Q^2)$ in lattice QCD}
We propose a new method to compute $\widehat\Pi(Q^2)$ without the problem of having to estimate $\Pi(Q^2=0)$ and that is calculable at any value of the virtuality $Q^2$. To this end, note the structure of the vacuum polarization tensor (\ref{eqn:tensorstruct}) implies:
      \be\la{eqn:tensorres}
      \Pi_{zz}(Q_0\hat e_0) = -Q_0^2 \Pi (Q_0^2)\quad.
      \ee
The time-momentum representation of the lattice vector correlation function is given by:
\be
G(x_0) \delta_{ik}= \int d^3x\, \langle J_i(x_0,\vec x) J_k(0) \rangle.
\ee
In combination with (\ref{eqn:tensorres}), the correlator $G(x_0)$ can be related to the hadronic vacuum polarization (\ref{eqn:fourcorr}) in the form $\Pi (Q_0^2)$ through
\begin{align}
\Pi (Q_0^2)= - \frac{\Pi_{zz}(Q_0)}{Q_0^2} 
=\frac{1}{Q_0^2} \int_{-\infty}^{\infty} dx_0 e^{iQ_0 x_0} G(x_0).
\end{align}
To obtain $\widehat{\Pi}(Q_0^2)=4\pi^2( \Pi(Q_0^2) - \Pi(0) )$ expand:
\begin{align}
\Pi(Q_0^2\rightarrow 0) = \frac{1}{Q_0^2}\int_{-\infty}^\infty dx_0\, G(x_0)
-\frac{1}{2}\int_{-\infty}^\infty dx_0\,x_0^2\,G(x_0)+ \mathcal{O}(Q_0^2) ...
\end{align}
Therefore the subtracted vacuum polarization $\widehat\Pi(Q^2)$ can be expressed as an integral over the current-current correlator $G(x_0)$:
\be\la{eqn:pihat}
\Pi(Q_0^2) - \Pi(0) = \int_0^{\infty} dx_0 G(x_0)  K_{\Pi}(Q_0^2,x_0^2)
\ee
where
\be\la{eqn:kernel}
K_{\Pi}(Q_0^2,x_0^2)=\Big[ x_0^2 - \frac{4}{Q_0^2}\sin^2(\frac{1}{2}Q_0x_0)\Big].
\ee
  With (\ref{eqn:pihat}) we obtain an integral representation that is convergent and gives $\widehat{\Pi}(Q_0^2)$ at any $Q^2$ without having to estimate $\Pi(Q^2=0)$.\\
Additionally, in this representation derivatives of $\Pi(Q^2)$ can easily be taken by changing the kernel $K_{\Pi}(Q_0^2,x_0^2)$ accordingly. The first derivative for example can be linked to the phenomenologically interesting Adler function 
\begin{align} 
 D(Q_0^2) \equiv 12\pi^2 Q_0^2 \frac{d\,\Pi}{dQ_0^2}
 = \frac{12\pi^2}{Q_0^2} \int_0^\infty dx_0 \, G(x_0) 
\left(2-2\cos(Q_0 x_0) - Q_0x_0\sin(Q_0x_0)\right).
\end{align}
In the limits $Q_0^2\rightarrow 0$ and $m_l\rightarrow 0$ the derivative of the Adler function gives $a_\mu^{\rm HLO}$ directly 
\begin{align} 
&D'(0)=\lim_{{Q^2\to0}}\frac{D(Q^2)}{Q^2}
= \pi^2 \int_0^\infty dx_0 \;x_0^4\, G(x_0)\\
\Rightarrow &\lim_{m_l\rightarrow 0} \frac{a_l^{HLO}}{m_l^2} = \frac{1}{9}\Big(\frac{\alpha}{\pi}\Big)^2 D'(0). \la{eqn:slopeadler}
\end{align} 
This entails the first derivative, i.e. the slope of $\Pi(Q^2)$ at the origin gives the bulk of the leading order hadronic contribution of the anomalous magnetic moments of leptons immediately and is directly accessible to the lattice using the proposed mixed representation method.
    
\section{Numerical Setup}

We test the procedure on dynamical gauge
configurations with two mass-degenerate quark flavors.  The gauge
action is the standard Wilson plaquette action \cite{Wilson:1974sk},
while the fermions were implemented via the O($a$) improved Wilson
discretization with non-perturbatively determined clover coefficient
$c_{\rm sw}$ \cite{Jansen:1998mx}.  The configurations
were generated within the CLS effort~\cite{CLS} and the used algorithms are
based on L\"uscher's DD-HMC package~\cite{CLScode}.  
We calculated
correlation functions using the same discretization and masses as in
the sea sector on a lattice of size $96\times 48^3$ (labeled F6
in~\cite{Capitani:2012gj}) with a lattice spacing of $a =
0.0631(21)$fm~\cite{Capitani:2011fg} and a pion mass of
$m_\pi=324$MeV, so that $m_\pi L = 5.0$.

In this lattice study we consider isospin-symmetric two-flavor QCD. 
The electromagnetic current is given by $j^{\gamma}_\mu = j^\rho_\mu + {\txts\frac{1}{3}} j^\omega_\mu$ with
\be
j_\mu^\rho \equiv \frac{1}{2} (\bar u \gamma_\mu u - \bar d \gamma_\mu d),\qquad
j_\mu^\omega \equiv \frac{1}{2} (\bar u\gamma_\mu u + \bar d\gamma_\mu d).
\ee
As such $j_\mu^\omega$ gives rise to both connected and disconnected diagrams for lattice QCD calculations. This kind of combination requires a calculation using all-to-all propagators in order to compute the Wick-disconnected quark loops.
Recently a very interesting relation was
obtained~\cite{Juttner:2009yb} using chiral perturbation theory stating that the relative strengths of the contributions of the
Wick-connected and the Wick-disconnected diagrams in the vacuum polarization tensor $\Pi_{\mu\nu}(Q^2)$ are 1:10. This result was rederived in \cite{Francis:2013fzp} based on general considerations. 

For this reason we choose to measure instead the isovector current $j^\rho_\mu$, as it does not contain disconnected diagrams by definition.
On the lattice we implemented the local-conserved isovector vector correlation function in the form:
\be\la{eq:Gdeflc}
G(x_0)\delta_{kl} =Z_V(g_0) G^{\rm bare}(x_0,g_0)\delta_{kl} 
= - a^3 Z_V(g_0)\sum_{\vec x} \langle J^c_k(x) J^l_\ell(0) \rangle,
\ee
\vspace{-2.4em}with:
\ba
J_\mu^l(x) &=& \bar q(x) \gamma_\mu q(x),
\\
\la{eq:Jdef}
J_\mu^c ( x ) &=& \frac{1}{2} \Big(\bar q ( x + a\hat\mu ) ( 1 + \gamma_\mu ) U_\mu^\dagger ( x ) q ( x ) 
 - \bar q ( x )( 1 - \gamma_\mu ) U_\mu ( x )q ( x + a\hat\mu ) \Big).
\ea
whereby we used the non-perturbative value of
$Z_V=0.750(5)$~\cite{DellaMorte:2005rd} to renormalize the lattice results.

 \section{Numerical Results}

\begin{figure}[t]  
\hspace{-1.4em}\includegraphics[width=0.54\linewidth]{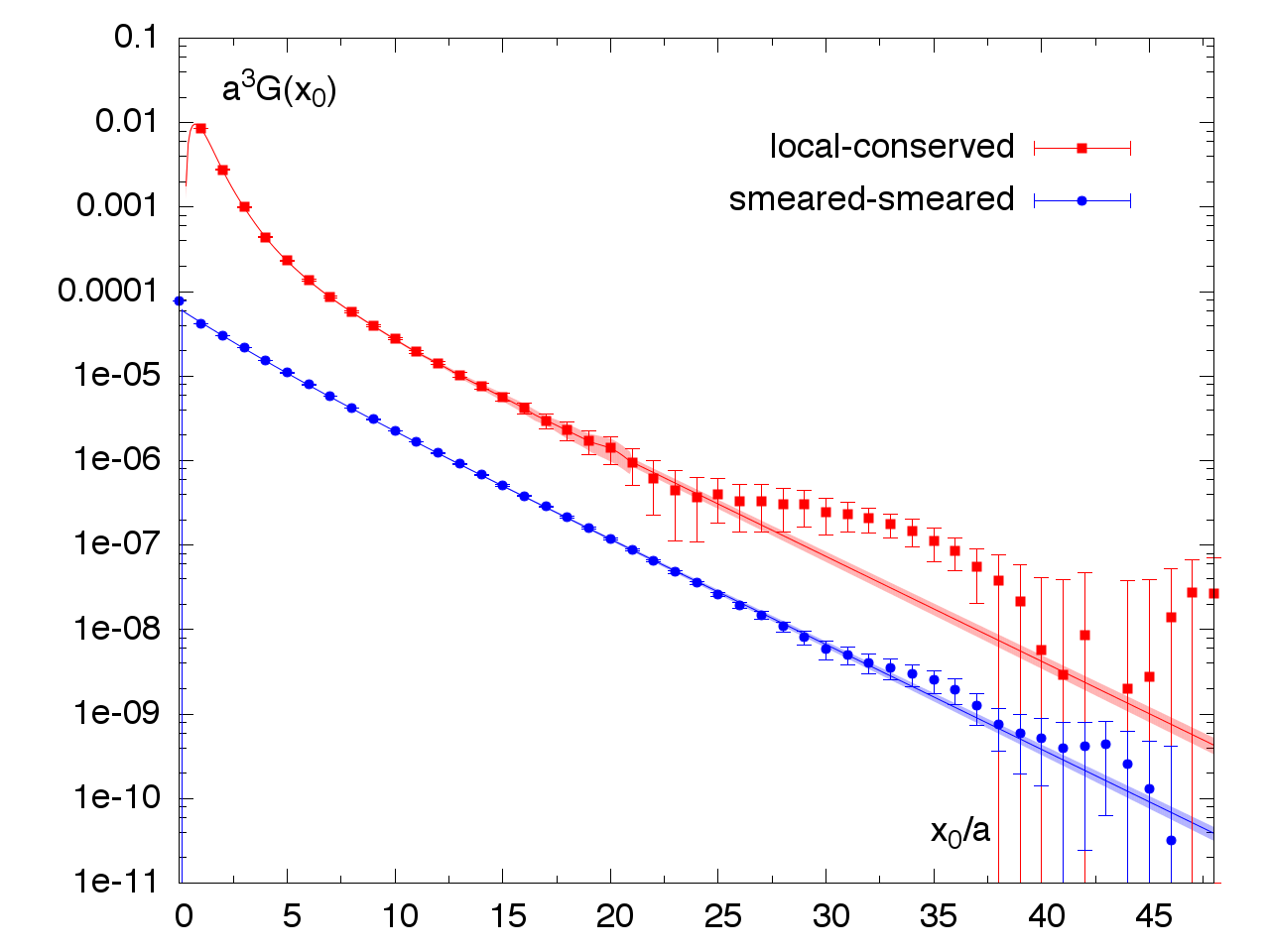} 
\hspace{-1em}\includegraphics[width=0.54\linewidth]{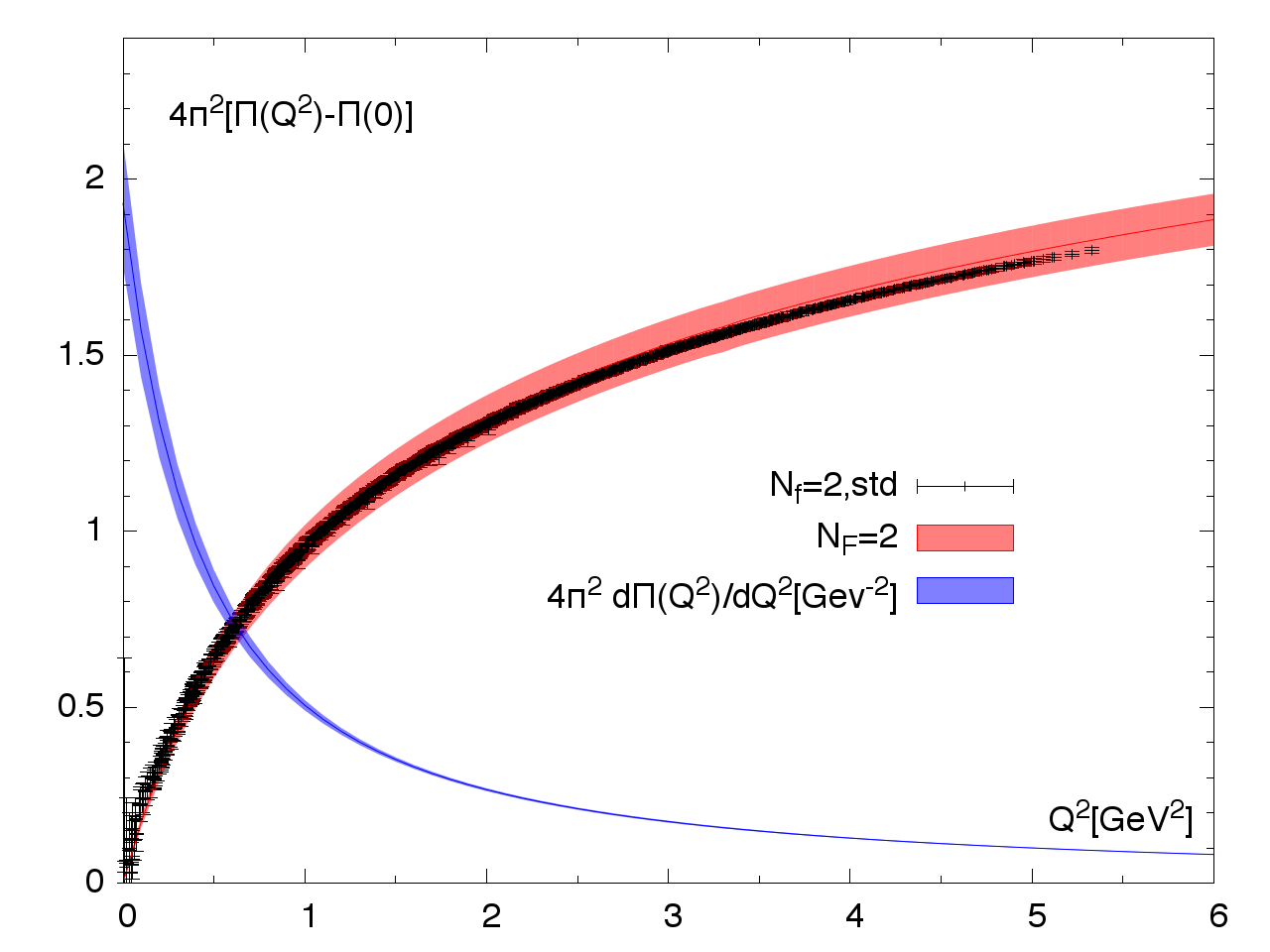} 
\caption{\it{Left: The local-conserved and smeared-smeared isovector vector correlation functions. 
The red shaded area shows the correlator entering 
the computation of $\widehat\Pi(Q^2)$, the blue shaded area correlator is used 
to fit the lowest lying mass for extrapolation to all time beyond $x_0\simeq T/4$.
Right: The subtracted vacuum polarization $\widehat\Pi(Q^2)$ and $d\widehat\Pi(Q^2)/dQ^2$ 
computed from the extended lattice correlator (left). 
The data shown in black were obtained using the momentum-space method 
on the same ensemble with comparable statistics \cite{DellaMorte:2012cf}.
}}
\label{fig:corr}
\end{figure}
  
 Based on (\ref{eqn:pihat}) our aim is the computation of $\widehat\Pi(Q^2)$ and its derivative from the lattice correlation function. This requires the integration of the correlator (\ref{eq:Gdeflc}) convoluted with the kernel (\ref{eqn:kernel}) over all time separations. On the lattice, however, only a finite number of points is available and a continuation of the lattice data to all time separations becomes necessary. 
At $m_\pi L=5.0$ and $m_\pi=324$MeV it is safe to assume the $\rho$-particle is still stable and dominates the exponential decay of the correlator at long times. Consequently we extend the local-conserved correlator by fitting the lattice data with an exponential that
decays with the lowest lying mass of the system
\be
G_{\rm Ansatz}(x_0)=\sum_{n=1}^2 |A_n|^2 e^{-m_n x_0}\quad \textrm{for}~~~x_0 \ll T/2.
\ee
Here, we fix the lowest lying mass by extracting it from a separate smeared-smeared correlation function~\cite{Capitani:2011fg}, as these results exhibit greater overlap with the ground state. Subsequently the mass parameter
determined in this way is passed to the fit of the local-conserved
correlator and the corresponding exponential is smoothly connected to
the lattice data by adjusting, i.e. fitting, $|A_1|^2$ to the data around $x_0=T/4$.

The resulting local-conserved correlation function is shown as the red
shaded band in Fig.\ref{fig:corr}(left), while the smeared-smeared result is shown in blue.
The error estimates were
obtained via a jackknife procedure. In the transition region from the
data dominated to the extrapolation dominated results the errors
increase for a small number of time steps. Nevertheless this procedure yields a stable result for the local-conserved time-momentum vector correlator with small errors. 

 
Using this correlator we can compute the subtracted vacuum polarization $\widehat\Pi(Q^2)$ and the Adler function $D(Q^2)$. In Fig.\ref{fig:corr}(right) we show the resulting $\widehat\Pi(Q^2)$ (red) and its slope $4\pi^2 d\Pi(Q^2)/dQ^2$ (blue). For comparison we also show the result of  $\widehat\Pi(Q^2)$ obtained on the same lattice using the standard method
\cite{DellaMorte:2012cf} with the same local-conserved discretization
and comparable statistics (black). In these data the number of available virtualities was significantly boosted using twisted-boundary conditions
\cite{deDivitiis:2004kq,Sachrajda:2004mi,Bedaque:2004ax}. Nevertheless the extrapolation to $\Pi(Q^2=0)$ is non-trivial and difficult to constrain as the signal deteriorates as $Q^2$ approaches zero. \\
Clearly, the results obtained using (\ref{eqn:pihat}) are very well
compatible with the standard method.  It should be noted that the
larger errors for large $Q^2$ only play a small role when computing
$a_\mu^{\rm HLO}$, as the large $Q^2$ region is highly suppressed in
the relevant integral. 
Turning to the slope of $\widehat\Pi(Q^2)$ we find the result exhibits small statistical
errors and the intercept at $Q^2=0$ can be determined relatively
precisely.  Reading off the intercept we find $D'(0)=3\widehat\Pi'(0) = 5.8(5){\rm GeV}^{-2}$. In principle this value can be used to constrain the determination of the functional form of $\Pi(Q^2)$ in the standard method or to estimate $\alpha_l^{HLO}$ in the limit where $m_l\rightarrow 0$.

\begin{figure}[t]  
\centering
\includegraphics[width=0.75\linewidth]{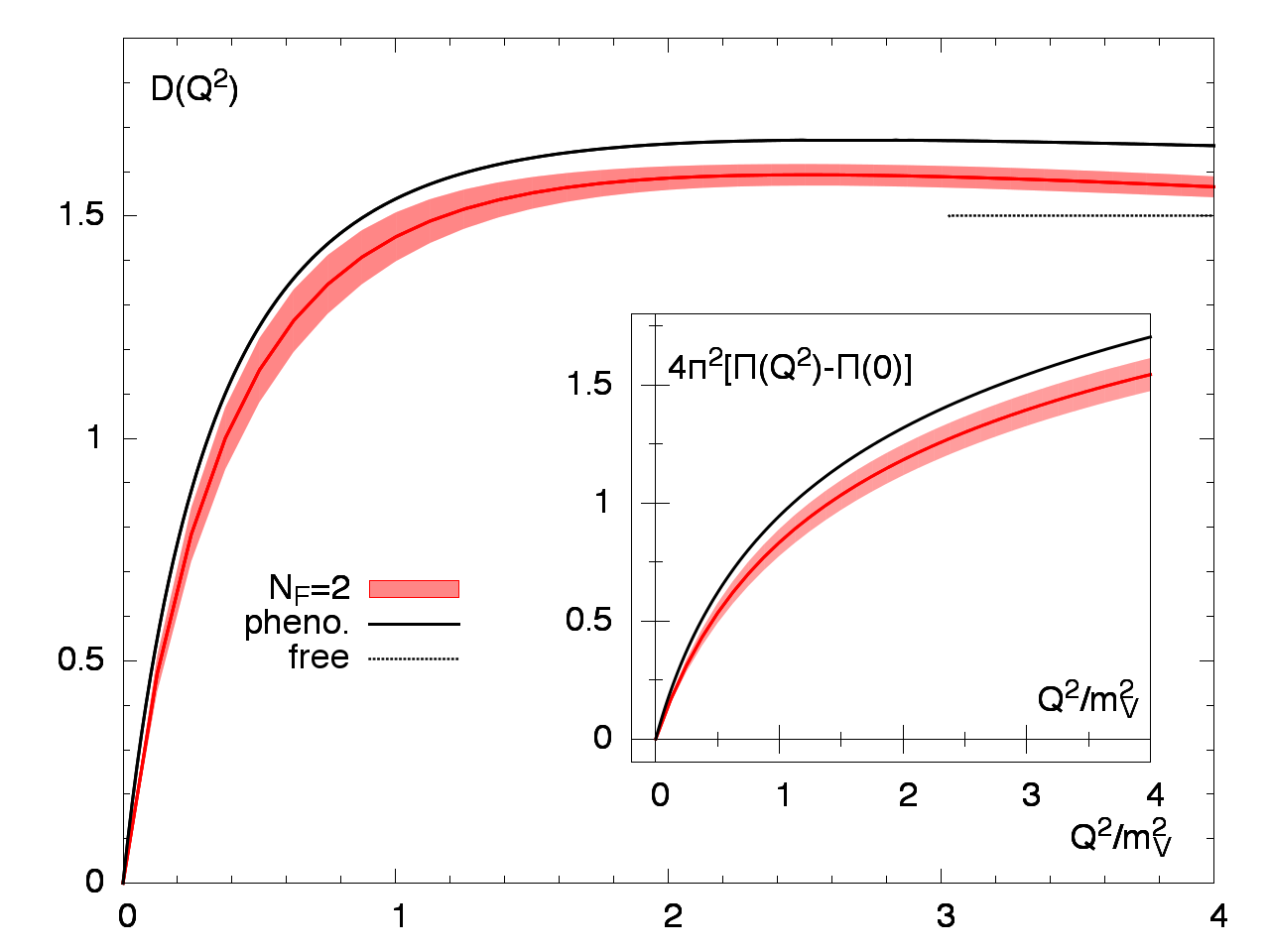} 
\caption{\it{The functions $\widehat\Pi(Q^2)$ and $D(Q^2)$ from our analysis and 
a phenomenological model \cite{Bernecker:2011gh}. 
The horizontal axis has been rescaled by the ground-state mass $m_1=894(2)$MeV
on the lattice and the physical $\rho$ meson mass ($770$MeV) respectively.
For reference  the free result $D(Q^2)=\frac{3}{2}$ is shown as a dotted line.}}
\label{fig:pihat2}
\end{figure}

In the next step we follow the approach of~\cite{gen4,Bernecker:2011gh}, which entails rescaling the horizontal axis using the vector meson mass with the aim of canceling the dominant chiral effects. As such one hopes to achieve an approximate scaling at small virtualities $Q^2$. Consequently the curves of the different quark mass scenarios should then fall on top of each other.
In this way we can compare the lattice results with those of phenomenology. In Fig.\ref{fig:pihat2} we show the Adler function and the vacuum polarization scaled in this way compared to the phenomenological model of \cite{Bernecker:2011gh} for the isovector channel. Clearly, the lattice data lies below the phenomenological curves. Additionally the intercept $D'(0)$ is seen to be roughly a factor 1.7 smaller than the value of the phenomenological model, given by $D'(0)=9.81(30){\rm GeV}^{-2}$. This might be due to the different spectral densities below the $\rho$ mass in the lattice and phenomenological cases \cite{Francis:2013fzp,Bernecker:2011gh}.

\section{Conclusion}

In conclusion, we implemented a new representation of the hadronic vacuum polarization, that does not require an extrapolation $Q^2\rightarrow 0$ and is available at any virtuality. In addition it enables the direct computation of the Adler function and the slope of $\widehat{\Pi}(Q^2)$, without additional approximation.
This method proposes a new way to systematically study and improve the results on the leading hadronic contributions to the anomalous magnetic moments of leptons using lattice QCD. It is based on the well known time-momentum representation of the vector lattice correlation function and enables the use of the sophisticated tools of spectroscopy to study for example the finite size and lattice discretization effects \cite{Francis:2013fzp}. 
The insights obtained in this framework will benefit the 
understanding of the systematic effects on the lattice QCD based value
of $a_\mu^{\rm HLO}$
and $\Delta \alpha(M_Z^2)$.
 
\vfill 
 
\acknowledgments
We are grateful to Michele Della Morte and Andreas J\"uttner for providing a code basis and to our colleagues within CLS for sharing the lattice ensemble used.  
We thank Georg von Hippel for discussions and for providing the smeared 
vector correlator~\cite{Capitani:2011fg}.
The correlation functions were computed 
on the dedicated QCD platform ``Wilson'' at the Institute for Nuclear Physics,
University of Mainz. 
This work was supported by the \emph{Center for Computational Sciences}
as part of the Rhineland-Palatinate Research Initiative.

\end{document}